\documentclass{iopjournal}

\usepackage{graphicx}
\usepackage{bm}
\usepackage{ragged2e}
\usepackage{lipsum}
\usepackage{amsmath}
\usepackage{amsthm}
\usepackage{mathrsfs}
\usepackage{amsfonts}
\usepackage{amssymb}
\usepackage{braket} 
\usepackage{multirow}
\usepackage{booktabs}
\usepackage{textcomp}
\usepackage{appendix}
\usepackage{cite}

\begin{document}

\articletype{Article type} 

\title{Unambiguous arbitrary high-dimensional Bell states analyzer via indefinite causal order}

\author{Jun-Hai Zhao$^{1}$\orcid{0009-0008-2193-8299}, Wen-Qiang Liu$^2$\orcid{0000-0001-9804-2716} and Hai-Rui Wei$^{1,*}$\orcid{0000-0001-7459-4161}}

\affil{$^1$School of Mathematics and Physics, University of Science and Technology Beijing, Beijing 100083, China}

\affil{$^2$Department of Mathematics and Physics, Shijiazhuang Tiedao University, Shijiazhuang 050043, China}

\affil{$^*$Author to whom any correspondence should be addressed.}

\email{hrwei@ustb.edu.cn}


\begin{abstract} \justifying

High-dimensional quantum systems greatly outperform their two-dimensional counterparts in channel capacity, quantum complexity and efficiency,  quantum communication security, etc.
Bell-state analyzer (BSA) is a crucial prerequisite for a number of quantum communication protocols.
We propose an approach for completely and deterministically distinguishing a set of arbitrary $d$-dimensional ($d \geq 3$) Bell states via indefinite causal order (ICO).
In previous schemes, bit and phase information are discriminated in succession. 
Exploiting the gravitational ICO as the sole resource, we propose some high-dimensional BSA schemes. 
Independent of the dimensions, a set of generalized Bell states are completely and deterministically discriminated by adjusting the form of the embedded local single-qudit gates within ICO switch and measuring each qudit in the $\{|0\rangle, |1\rangle, \cdots, |d-1\rangle\}$ basis.
Notably, in our high-dimensional BSA process, the indefinite causal structure is not consumed. 
Hence a completely nondestructive high-dimensional BSA can be achieved by iterating the indefinite causal structure process for two rounds.

\end{abstract}

\section{Introduction}  \label{sec1}

\justifying

Quantum entanglement is essential for many quantum information processing (QIP) tasks, such as quantum key distribution \cite{QKD1,QKD2}, quantum dense coding \cite{Q-dense-code-1,Q-dense-code-2}, quantum teleportation \cite{Q-teleport1,Q-teleport2}, quantum secret sharing \cite{QSS1,QSS2}, quantum entanglement swapping \cite{Q-entanglement-swap1,Q-entanglement-swap2}, and quantum secure direct communication. An essential step in these protocols is the Bell-state analyzer (BSA) \cite{BSA1,BSA2}, which is defined as an approach to distinguish all four of the Bell states. 
Complete and deterministic BSA is a significant prerequisite for many quantum communication tasks.
Nowadays, BSA protocols have been proposed in a number of two-dimensional physical systems \cite{photon-BSA1,atom-BSA,hyper-BSA-1}.

Implementing a complete and deterministic Bell-state analyzer (BSA) still faces fundamental challenges when restricted to local operations and classical communication (LOCC). 
The LOCC-based protocols can perfectly discriminate any two orthogonal multipartite pure states \cite{theroy-LOCC-1, theroy-LOCC-2}. Unfortunately, this LOCC-based perfect discriminating capability is fundamentally limited when focused on a set of three (or more) entangled states. The presence of entanglement poses the key obstacle, and research confirms that the number of locally distinguishable states is constrained by the total dimension of the system over the average entanglement of the state \cite{theroy-LOCC-3, theroy-LOCC-4}.

BSA was first proposed by Kwiat and Weinfurter \cite{BSA1} in 1998, and it relies only on linear optical elements. Subsequently, tremendous progress has been made both in theory and experiment. Polarization BSA has been experimentally demonstrated using local linear optics \cite{photon-BSA-dense-coding,photon-BSA1,photon-BSA2}, and the optimal efficiency of these approaches is 50\%. One strategy adopted to overcome this inherent probabilistic nature is to employ ``buses'', ranging from atoms \cite{atom-BSA}, cross-Kerr media \cite{photon-Kerr}, and artificial atoms \cite{Artificial-Atom-BSA1} to superimpose different operations \cite{gravity-BSA}. The second strategy is to introduce additional hyper-entangled Bell pairs \cite{hyper-assisted-BSA}. The third strategy is to exploit additional degrees of freedom \cite{DOF-BSA1,DOF-BSA2}, utilize conditioned detection \cite{condition-BSA}, or enlarge the Hilbert space \cite{high-OAM-BSA}. The existing BSAs are mainly focused on 2-dimensional systems, while their high-dimensional scenarios remain largely unexplored.

In traditional QIP models, quantum operations are applied in a fixed sequential order. However, quantum mechanics also permits the scenarios in which two or more events take place in an indefinite order \cite{ICO}. This ``superimpose different operations'' is called indefinite causal order (ICO), and such quantum-controlled order of events has been experimentally confirmed \cite{ICO-confirm2,ICO-confirm3,ICO-independent,ICO-probabilistic}. It has been demonstrated that ICO can offer notable operational advantages over their classical fixed scenarios in various specific QIP tasks, including quantum channel discrimination \cite{causal-model}, promise problems \cite{promise-problem}, communication complexity tasks \cite{communication-complexity1,communication-complexity2}, quantum communication \cite{communication}, quantum metrology \cite{ICOMetrology1,ICOMetrology2}, quantum thermodynamics \cite{thermodynamic1,thermodynamic2}, quantum algorithms \cite{algorithm}, noise mitigation \cite{mitgation}, quantum key distribution \cite{ICO-QKD}, and entanglement generation \cite{generation} and distillation \cite{distillation}.

In this work, we propose a more elegant complete and deterministic BSA in an arbitrary high-dimensional system. 
Exploiting ICO as the sole resource, we first propose an unambiguous BSA scheme in 3-dimensional quantum system. Then, we extend our ICO-based approach to a 4-dimensional scenario and further extend it to arbitrary $d$-dimensional scenarios. Subsequently, by leveraging the quantum superposition of different space-time geometries, we design a set of high-dimensional ICO gravitational switches to perfectly distinguish the corresponding set of high-dimensional generalized Bell states.

The scheme  we designed for the high-dimensional BSA has the following characteristics. 
First, single-partite unitary operations are essential for our protocol. A perfect high-dimensional BSA is completed by adjusting the forms of the single-partite operations. Therefore, our scheme bypasses the need for nonlocal operations, material platforms, or additional entangled pairs, which are necessary for existing BSA protocols. Second, only local shift gates involved in the ICO switch are introduced, and these simple single-qudit gates are experimentally feasible.
Third, non-destructive high-dimensional BSA can be achieved by iterating the ICO switch for two rounds, as the ICO is not consumed. Fourth, after one switch process and a projection measurement of the Bell state, the high-dimensional Bell states can be completely distinguished. In previous two-step schemes, bit and phase information are distinguished in succession.

The organization of this paper is as follows. In Sec. \ref{sec2}, we first present the limitations of high-dimensional BSA within LOCC framework. 
Subsequently, we propose a scheme to completely distinguish the nine qutrit Bell states via ICO quantum 3-switch. Finally, we generalize our method to arbitrary 4-dimensional and further to arbitrary $d$-dimensional scenarios. 
In Sec. \ref{sec3}, we first propose an architecture to implement space-time ICO quantum 3-switch, where the space-time ICO arises from the spatial superposition of a massive gravitating body. 
Subsequently, we extend our approach to the gravitational ICO quantum 4-switch and further to gravitational ICO quantum $d$-switch scenarios.
We provide a discussion and summarize our results in Sec. \ref{sec4}.


\section{Complete and deterministic BSA in $d$-dimensional system via ICO}\label{sec2}

\subsection{The limitation of the traditional LOCC-based BSA}\label{sec21}

We consider the two agents Alice and Bob who share a bipartite quantum system. Each spatially separated party can locally perform quantum operations on their own subsystem and communicates their respective classical information with each other. This LOCC is denoted by $\mathcal{O}_L$. Obviously, set $\mathcal{O}_L$ forms a strict subset of the set of all quantum operations (denoted by $\mathcal{O}$) performed on the arbitrary bipartite quantum state, i.e., $\mathcal{O}_L \subset \mathcal{O}$. Operations which cannot be implemented by solely using LOCC are called nonlocal operations (denoted by $\mathcal{O}_{NL}$).

In bipartite $d$-dimensional system, there are $d^2$ generalized Bell states, and  
the aim of this paper is to distinguish this set of the orthogonal generalized Bell states completely and deterministically. 
It is known that it is impossible to perfectly distinguish all $d^2$ generalized Bell states by using only standard LOCC. 
Early in 2006, Hayashi et al. \cite{theroy-LOCC-3} provided a strict quantitative upper bound on the number $N$ of pure states that can be perfectly distinguished by LOCC, i.e., 
%
\begin{eqnarray}\label{eq1}
N(\text{Bi}) \le d_1 d_2 / (\sum_i \alpha_i)^2.
\end{eqnarray}
Here the $N$ pure states have the same entanglement (Bi). 
$d_1$ and $d_2$ are the dimensions of the two subsystems, respectively. 
$\alpha_i$ is the Schmidt coefficient for each pure states contained in the set.

Substituting $d_1 = d_2 = d$  and $(\sum_i \alpha_i)^2 = d$ into Eq. \eqref{eq1}, we can obtain 
\begin{eqnarray}\label{eq2}
N \le (d \cdot d) / d = d.
\end{eqnarray}
This indicates that it is impossible to distinguish more than $d$ generalized Bell states in $d$-dimensional system by two spatially separated parties using standard LOCC. 
It is thus desirable to seek alternative efficient scheme for completely implementing high-dimensional BSA in a deterministic way.
In the following, we will access perfect BSA with assistance of ICO.

\subsection{Deterministic Bell-state analyzer in 3-dimensional system via ICO 3-switch}\label{sec22}

The ICO quantum switch $\mathcal{S}$, the simplest and most common form of ICO, applies two or more events to the target system in a temporal order coherently controlled by an additional control system \cite{switch}. 
Now, let us detail the ICO quantum 3-switch operation.

The ICO quantum 3-switch $\mathcal{S}^3$ serves as a universal resource to map the different orthogonal basis states $|k\rangle_c$ of the additional control system to different local qutrit (i.e., three-state or three-level quantum system) operations $U_k^3 \otimes V_k^3$ on the target system. Here the subscript $k=1,2,3$, and the superscript 3 denotes the 3-dimensional system.  
For clarity and concise exposition of the working principles of $\mathcal{S}^3$, the control system of $\mathcal{S}^3$ is initially prepared in the superposition state $|\mathcal{F}_0\rangle^c$. Here
\begin{eqnarray}\label{eq3}
 |\mathcal{F}_0\rangle^c = \frac{1}{\sqrt 3 } \left( | 0 \rangle^c + | 1 \rangle^c + | 2 \rangle^c \right).
\end{eqnarray}
The two target subsystems Alice and Bob are initially prepared in the product state, $|\Psi\rangle_A^t \otimes |\Psi\rangle_B^t \in \mathbb{C}^3 \otimes \mathbb{C}^3$.
The operation of $\mathcal{S}^3$ is given by
\begin{equation}\label{eq4Cite}
\begin{split}
\mathcal{S}^3 |\mathcal{F}_0\rangle^c |\Psi\rangle_A^t |\Psi\rangle_B^t 
=& \frac{1}{\sqrt 3 } \big(|0\rangle^c \otimes U_1^3 |\Psi\rangle_A^t \otimes V_1^3 |\Psi\rangle_B^t  + |1\rangle^c \otimes U_2^3 |\Psi\rangle_A^t \otimes V_2^3 |\Psi\rangle_B^t \\
                       & + |2\rangle^c \otimes U_3^3 |\Psi\rangle_A^t \otimes V_3^3 |\Psi\rangle_B^t \big) \\
=& \frac{1}{3 } |\mathcal{F}_0\rangle^c\big[(U_1^3 \otimes V_1^3) + (U_2^3 \otimes V_2^3)                    + (U_3^3 \otimes V_3^3) \big]|\Psi \rangle_A^t|\Psi \rangle_B^t \\
 & +\frac{1}{3 }|\mathcal{F}_1\rangle^c\big[(U_1^3 \otimes V_1^3) + e^{i\frac{4\pi}{3}}(U_2^3 \otimes V_2^3) + e^{i\frac{2\pi}{3}}(U_3^3 \otimes V_3^3)\big]|\Psi \rangle _A^t|\Psi \rangle _B^t \\
 & +\frac{1}{3 }|\mathcal{F}_2\rangle^c\big[(U_1^3 \otimes V_1^3) + e^{i\frac{2\pi}{3}}(U_2^3 \otimes V_2^3) + e^{i\frac{4\pi}{3}}(U_3^3 \otimes V_3^3)\big]|\Psi \rangle _A^t|\Psi \rangle _B^t.
\end{split}
\end{equation}

where
\begin{eqnarray}   \label{eq26}
|\mathcal{F}_0\rangle  = \frac{1}{{\sqrt 3 }}{(|0\rangle^c} + |1{\rangle^c} + |2{\rangle^c}),
\end{eqnarray}
\begin{eqnarray}   \label{eq27}
|\mathcal{F}_1\rangle  = \frac{1}{{\sqrt 3 }}{(|0\rangle^c} + {e^{\frac{{2\pi {\rm{i}}}}{3}}}|1{\rangle^c} + {e^{\frac{{4\pi {\rm{i}}}}{3}}}|2{\rangle^c}),
\end{eqnarray}
\begin{eqnarray}   \label{eq28}
|\mathcal{F}_2\rangle  = \frac{1}{{\sqrt 3 }}{(|0\rangle^c} + {e^{\frac{{4\pi {\rm{i}}}}{3}}}|1{\rangle^c} + {e^{\frac{{2\pi {\rm{i}}}}{3}}}|2{\rangle^c}).
\end{eqnarray}

Based on Eq. \eqref{eq4Cite}, we can see that depending on the output states $\{|\mathcal{F}_0\rangle$, $|\mathcal{F}_1\rangle$, $|\mathcal{F}_2\rangle\}$, 
the output target state is an entangled state by suitable choice of local qutrit operations $U_k^3$ and $V_k^3$ ($k=1,2,3$).  
Hence, the ICO switch $\mathcal{S}^3$ described by Eq. \eqref{eq4Cite} serves as a sole resource and has the potential to  completely discriminate nine qutrit generalized Bell states $\{|\Psi_{i,j}\rangle\}$ with $i,j \in \{0,1,2\}$. Here
\begin{eqnarray}   \label{eq5}
    |\Psi_{0,0}\rangle_{AB} = \frac{1}{\sqrt{3}} (|00\rangle + |11\rangle + |22\rangle)_{AB},
\end{eqnarray}
\begin{eqnarray}   \label{eq6}
    |\Psi_{0,1}\rangle_{AB} = \frac{1}{\sqrt{3}} (|01\rangle + |12\rangle + |20\rangle)_{AB},
\end{eqnarray}
\begin{eqnarray}   \label{eq7}
    |\Psi_{0,2}\rangle_{AB} = \frac{1}{\sqrt{3}} (|02\rangle + |10\rangle + |21\rangle)_{AB},
\end{eqnarray}
\begin{eqnarray}   \label{eq8}
    |\Psi_{1,0}\rangle_{AB} = \frac{1}{\sqrt{3}} (|00\rangle + e^{\frac{2\pi\text{i}}{3}} |11\rangle + e^{\frac{4\pi\text{i}}{3}} |22\rangle)_{AB},
\end{eqnarray}
\begin{eqnarray}   \label{eq9}
    |\Psi_{1,1}\rangle_{AB} = \frac{1}{\sqrt{3}} (|01\rangle + e^{\frac{2\pi\text{i}}{3}} |12\rangle + e^{\frac{4\pi\text{i}}{3}} |20\rangle)_{AB},
\end{eqnarray}
\begin{eqnarray}   \label{eq10}
    |\Psi_{1,2}\rangle_{AB} = \frac{1}{\sqrt{3}} (|02\rangle + e^{\frac{2\pi\text{i}}{3}} |10\rangle + e^{\frac{4\pi\text{i}}{3}} |21\rangle)_{AB},
\end{eqnarray}
\begin{eqnarray}   \label{eq11}
    |\Psi_{2,0}\rangle_{AB} = \frac{1}{\sqrt{3}} (|00\rangle + e^{\frac{4\pi\text{i}}{3}} |11\rangle + e^{\frac{2\pi\text{i}}{3}} |22\rangle)_{AB},
\end{eqnarray}
\begin{eqnarray}   \label{eq12}
    |\Psi_{2,1}\rangle_{AB} = \frac{1}{\sqrt{3}} (|01\rangle + e^{\frac{4\pi\text{i}}{3}} |12\rangle + e^{\frac{2\pi\text{i}}{3}} |20\rangle)_{AB},
\end{eqnarray}
\begin{eqnarray}   \label{eq13}
    |\Psi_{2,2}\rangle_{AB} = \frac{1}{\sqrt{3}} (|02\rangle + e^{\frac{4\pi\text{i}}{3}} |10\rangle + e^{\frac{2\pi\text{i}}{3}} |21\rangle)_{AB}.
\end{eqnarray}
The embedded local operations $U_k^3$ and $V_k^3$ ($k=1,2,3$) within $\mathcal{S}^3$ given in Eq. \eqref{eq4Cite} are adjusted as the following exponentiation functions
\begin{eqnarray}\label{eq14}
 U_1^3 = (U_{\text{shift}}^3)^0, \quad U_2^3 = (U_{\text{shift}}^3)^1, \quad   U_3^3 = (U_{\text{shift}}^3)^2, 
\end{eqnarray}
\begin{eqnarray}\label{eq15}
 V_1^3 = (U_{\text{shift}}^3)^1, \quad  V_2^3 = (U_{\text{shift}}^3)^2, \quad  V_3^3 = (U_{\text{shift}}^3)^3.
\end{eqnarray}
Here
\begin{eqnarray}\label{eq16}
U_{\text{shift}}^3 = 
    \begin{pmatrix}
    0 & 0 & 1 \\
    1 & 0 & 0 \\
    0 & 1 & 0
\end{pmatrix}.
\end{eqnarray}
The exponentiation function $(U_{\text{shift}}^3)^0$ is a $3 \times 3$ identity matrix, 
$(U_{\text{shift}}^3)^2 = U_{\text{shift}}^3 \cdot U_{\text{shift}}^3$, and 
$(U_{\text{shift}}^3)^3 = U_{\text{shift}}^3 \cdot U_{\text{shift}}^3 \cdot U_{\text{shift}}^3$.

Based on Eq. \eqref{eq14} - Eq. \eqref{eq16}, we can see that the conversions of $\{|\Psi_{i,j}\rangle_{AB}\}$ induced by $\mathcal{S}^3$ are given by as
\begin{eqnarray}   \label{eq17}
 \begin{split}
    \mathcal{S}^3  |\mathcal{F}_0\rangle^c |\Psi_{0,0}\rangle_{AB} = |\mathcal{F}_0\rangle^c |\Psi_{0,1}\rangle_{AB},
 \end{split}
\end{eqnarray}
\begin{eqnarray}   \label{eq18}
 \begin{split}
    \mathcal{S}^3|\mathcal{F}_0\rangle^c|\Psi_{0,1}\rangle_{AB}  = |\mathcal{F}_0\rangle^c |\Psi_{0,2}\rangle_{AB},
 \end{split}
\end{eqnarray}
\begin{eqnarray}   \label{eq19}
 \begin{split}
    \mathcal{S}^3|\mathcal{F}_0\rangle^c|\Psi_{0,2}\rangle_{AB}  = |\mathcal{F}_0\rangle^c |\Psi_{0,0}\rangle_{AB},
 \end{split}
\end{eqnarray}
\begin{eqnarray}   \label{eq20}
 \begin{split}
    \mathcal{S}^3 |\mathcal{F}_0\rangle^c |\Psi_{1,0}\rangle_{AB} = |\mathcal{F}_2\rangle^c |\Psi_{1,1}\rangle_{AB},
 \end{split}
\end{eqnarray}
\begin{eqnarray}   \label{eq21}
 \begin{split}
    \mathcal{S}^3 |\mathcal{F}_0\rangle^c |\Psi_{1,1}\rangle_{AB} = |\mathcal{F}_2\rangle^c |\Psi_{1,2}\rangle_{AB},
 \end{split}
\end{eqnarray}
\begin{eqnarray}   \label{eq22}
 \begin{split}
    \mathcal{S}^3 |\mathcal{F}_0\rangle^c |\Psi_{1,2}\rangle_{AB} = |\mathcal{F}_2\rangle^c |\Psi_{1,0}\rangle_{AB},
 \end{split}
\end{eqnarray}
\begin{eqnarray}   \label{eq23}
 \begin{split}
  \mathcal{S}^3 |\mathcal{F}_0\rangle^c |\Psi_{2,0}\rangle_{AB} = |\mathcal{F}_1\rangle^c |\Psi_{2,1}\rangle_{AB},
 \end{split}
\end{eqnarray}
\begin{eqnarray}   \label{eq24}
 \begin{split}
  \mathcal{S}^3 |\mathcal{F}_0\rangle^c |\Psi_{2,1}\rangle_{AB} = |\mathcal{F}_1\rangle^c |\Psi_{2,2}\rangle_{AB},
 \end{split}
\end{eqnarray}
\begin{eqnarray}   \label{eq25}
 \begin{split}
  \mathcal{S}^3 |\mathcal{F}_0\rangle^c |\Psi_{2,2}\rangle_{AB} = |\mathcal{F}_1\rangle^c |\Psi_{2,0}\rangle_{AB},
 \end{split}
\end{eqnarray}

Based on Eq. \eqref{eq17} - Eq. \eqref{eq25}, we can see that according to the information of the control system, the nine Bell states $\{|\Psi_{i,j}\rangle_{AB}\}$ can be divided into three groups:
$\{ |\Psi_{0,0}\rangle_{AB}$, $|\Psi_{0,1}\rangle_{AB}$, $|\Psi_{0,2}\rangle_{AB} \}$ corresponding to $|\mathcal{F}_0\rangle^c$;
$\{ |\Psi_{1,0}\rangle_{AB}$, $|\Psi_{1,1}\rangle_{AB}$, $|\Psi_{1,2}\rangle_{AB} \}$ corresponding to $|\mathcal{F}_2\rangle^c$; and
$\{ |\Psi_{2,0}\rangle_{AB}$, $|\Psi_{2,1}\rangle_{AB}$, $|\Psi_{2,2}\rangle_{AB} \}$ corresponding to $|\mathcal{F}_1\rangle^c$.

The next task is only to distinguish the different relative value in each group and it can be accomplished by the projection measurement of each agent in the basis $\{|0\rangle, |1\rangle, |2\rangle\}$, see Table \ref{Tab1}.
Therefore, the nine 2-qutrit Bell states can be completely and deterministically distinguished by using the ICO quantum 3-switch $\mathcal{S}^3$.

\begin{table*}[htbp]
\centering
\caption{The correspondences between the outcome of the quantum 3-switch $\mathcal{S}^3$, qutrit information, and the state transition for discriminating nine Bell states.} \label{Tab1}
\begin{tabular}{cccc}
\hline\hline

 Outcome of 3-switch $\mathcal{S}^3$ & Qutrit information of ($j_A, j_B$) &  Output Bell state       &  Input Bell state \\

\hline

\multirow{3}{*}{$\ket{\mathcal{F}_0}^c$}       
                                     &  $j_B = (j_A + 1)\bmod 3$         &  $\ket{\Psi_{0,1}}_{AB}$  &  $\ket{\Psi_{0,0}}_{AB}$ \\
                                     &  $j_B = (j_A + 2)\bmod 3$         &  $\ket{\Psi_{0,2}}_{AB}$  &  $\ket{\Psi_{0,1}}_{AB}$ \\
                                     &  $j_B = j_A$                      &  $\ket{\Psi_{0,0}}_{AB}$  &  $\ket{\Psi_{0,2}}_{AB}$ \\
 
\addlinespace[0.8em]

\multirow{3}{*}{$\ket{\mathcal{F}_2}^c$}      
                                     & $j_B = (j_A + 1)\bmod 3$          &  $\ket{\Psi_{1,1}}_{AB}$  &  $\ket{\Psi_{1,0}}_{AB}$ \\
                                     & $j_B = (j_A + 2)\bmod 3$          &  $\ket{\Psi_{1,2}}_{AB}$  &  $\ket{\Psi_{1,1}}_{AB}$ \\
                                     & $j_B = j_A$                       &  $\ket{\Psi_{1,0}}_{AB}$  &  $\ket{\Psi_{1,2}}_{AB}$ \\
 
\addlinespace[0.8em]

\multirow{3}{*}{$\ket{\mathcal{F}_1}^c$}       
                                     & $j_B = (j_A + 1)\bmod 3$         &  $\ket{\Psi_{2,1}}_{AB}$   &  $\ket{\Psi_{2,0}}_{AB}$ \\
                                     & $j_B = (j_A + 2)\bmod 3$         &  $\ket{\Psi_{2,2}}_{AB}$   &  $\ket{\Psi_{2,1}}_{AB}$ \\
                                     & $j_B = j_A$                      &  $\ket{\Psi_{2,0}}_{AB}$   &  $\ket{\Psi_{2,2}}_{AB}$ \\

\hline\hline
\end{tabular}
\end{table*}

\subsection{Deterministic Bell-state analyzer in 4-dimensional system via ICO 4-switch}\label{sec23}

The operation of the quantum ICO quantum 4-switch $\mathcal{S}^4$ is given by
\begin{eqnarray}\label{eq29Cite}
\begin{split}
\mathcal{S}^4 |\mathcal{H}_0\rangle^c |\Psi\rangle_A^t |\Psi\rangle_B^t = &\; \frac{1}{\sqrt{4}}  
    (|0\rangle^c \otimes U_1^4 |\Psi\rangle_A^t \otimes V_1^4 |\Psi\rangle_B^t 
   + |1\rangle^c \otimes U_2^4 |\Psi\rangle_A^t \otimes V_2^4 |\Psi\rangle_B^t \\
 & + |2\rangle^c \otimes U_3^4 |\Psi\rangle_A^t \otimes V_3^4 |\Psi\rangle_B^t 
   + |3\rangle^c \otimes U_4^4 |\Psi\rangle_A^t \otimes V_4^4 |\Psi\rangle_B^t).
 \end{split}
\end{eqnarray}
Here and afterwards,
\begin{eqnarray}   \label{eq30}
    |\mathcal{H}_0\rangle^c = \frac{1}{2}(|0\rangle + |1\rangle + |2\rangle + |3\rangle),
\end{eqnarray}%
\begin{eqnarray}   \label{eq31}
    |\mathcal{H}_1\rangle^c = \frac{1}{2}(|0\rangle + \text{i}|1\rangle - |2\rangle - \text{i}|3\rangle),
\end{eqnarray}%
\begin{eqnarray}   \label{eq32}
    |\mathcal{H}_2\rangle^c = \frac{1}{2}(|0\rangle - |1\rangle + |2\rangle - |3\rangle),
\end{eqnarray}
\begin{eqnarray}   \label{eq33}
    |\mathcal{H}_3\rangle^c = \frac{1}{2}(|0\rangle - \text{i}|1\rangle - |2\rangle + \text{i}|3\rangle).
\end{eqnarray}

The local ququart (i.e., 4-level or 4-state quantum system) operations $U_k^4$ and $V_k^4$ with $k=1,2,3,4$  given in Eq. \eqref{eq29Cite}  are set as
\begin{eqnarray}\label{eq34}
 U_1^4 = (U_{\text{shift}}^4)^0,\quad 
 U_2^4 = (U_{\text{shift}}^4)^1,\quad
 U_3^4 = (U_{\text{shift}}^4)^2,\quad
 U_4^4 = (U_{\text{shift}}^4)^3,  
\end{eqnarray}
\begin{eqnarray}\label{eq35}
 V_1^4 = (U_{\text{shift}}^4)^1,\quad  
 V_2^4 = (U_{\text{shift}}^4)^2,\quad
 V_3^4 = (U_{\text{shift}}^4)^3,\quad
 V_4^4 = (U_{\text{shift}}^4)^4,
\end{eqnarray}
where
\begin{eqnarray}  \label{eq36}
U_{\text{shift}}^4 = 
       \begin{pmatrix} 
       0 & 0 & 0 & 1 \\
       1 & 0 & 0 & 0 \\ 
       0 & 1 & 0 & 0 \\
       0 & 0 & 1 & 0
       \end{pmatrix}. 
\end{eqnarray} 
The sixteen generalized 2-ququart  Bell states $\{\ket{\psi_{i,j}}\}$ with $i,j=0,1,2,3$ can be expressed as 
\begin{eqnarray}   \label{eq37}
 \begin{split}
    \ket{\psi_{0,0}}_{AB} = \frac{1}{2} (\ket{00} + \ket{11} + \ket{22} + \ket{33})_{AB},
 \end{split}
\end{eqnarray}
\begin{eqnarray}   \label{eq38}
 \begin{split}
    \ket{\psi_{0,1}}_{AB} = \frac{1}{2} (\ket{01} + \ket{12} + \ket{23} + \ket{30})_{AB},
 \end{split}
\end{eqnarray}
\begin{eqnarray}   \label{eq39}
 \begin{split}
    \ket{\psi_{0,2}}_{AB} = \frac{1}{2} (\ket{02} + \ket{13} + \ket{20} + \ket{31})_{AB},
 \end{split}
\end{eqnarray}
\begin{eqnarray}   \label{eq40}
 \begin{split}
    \ket{\psi_{0,3}}_{AB} = \frac{1}{2} (\ket{03} + \ket{10} + \ket{21} + \ket{32})_{AB},
 \end{split}
\end{eqnarray}
\begin{eqnarray}   \label{eq41}
 \begin{split}
    \ket{\psi_{1,0}}_{AB} = \frac{1}{2} (\ket{00} + \text{i}\ket{11} - \ket{22} - \text{i}\ket{33})_{AB},
 \end{split}
\end{eqnarray}
\begin{eqnarray}   \label{eq42}
 \begin{split}
    \ket{\psi_{1,1}}_{AB} = \frac{1}{2} (\ket{01} + \text{i}\ket{12} - \ket{23} - \text{i}\ket{30})_{AB},
 \end{split}
\end{eqnarray}
\begin{eqnarray}   \label{eq42}
 \begin{split}
    \ket{\psi_{1,2}}_{AB} = \frac{1}{2} (\ket{02} + \text{i}\ket{13} - \ket{20} - \text{i}\ket{31})_{AB},
 \end{split}
\end{eqnarray}
\begin{eqnarray}   \label{eq44}
 \begin{split}
    \ket{\psi_{1,3}}_{AB} = \frac{1}{2} (\ket{03} + \text{i}\ket{10} - \ket{21} - \text{i}\ket{32})_{AB},
 \end{split}
\end{eqnarray}
\begin{eqnarray}   \label{eq45}
 \begin{split}
    \ket{\psi_{2,0}}_{AB} = \frac{1}{2} (\ket{00} - \ket{11} + \ket{22} - \ket{33})_{AB},
 \end{split}
\end{eqnarray}
\begin{eqnarray}   \label{eq46}
 \begin{split}
    \ket{\psi_{2,1}}_{AB} = \frac{1}{2} (\ket{01} - \ket{12} + \ket{23} - \ket{30})_{AB},
 \end{split}
\end{eqnarray}
\begin{eqnarray}   \label{eq47}
 \begin{split}
    \ket{\psi_{2,2}}_{AB} = \frac{1}{2} (\ket{02} - \ket{13} + \ket{20} - \ket{31})_{AB},
 \end{split}
\end{eqnarray}
\begin{eqnarray}   \label{eq48}
 \begin{split}
    \ket{\psi_{2,3}}_{AB} = \frac{1}{2} (\ket{03} - \ket{10} + \ket{21} - \ket{32})_{AB},
 \end{split}
\end{eqnarray}
\begin{eqnarray}   \label{eq49}
 \begin{split}
    \ket{\psi_{3,0}}_{AB} = \frac{1}{2} (\ket{00} - \text{i}\ket{11} - \ket{22} + \text{i}\ket{33})_{AB},
 \end{split}
\end{eqnarray}
\begin{eqnarray}   \label{eq50}
 \begin{split}
    \ket{\psi_{3,1}}_{AB} = \frac{1}{2} (\ket{01} - \text{i}\ket{12} - \ket{23} + \text{i}\ket{30})_{AB},
 \end{split}
\end{eqnarray}
\begin{eqnarray}   \label{eq51}
 \begin{split}
    \ket{\psi_{3,2}}_{AB} = \frac{1}{2} (\ket{02} - \text{i}\ket{13} - \ket{20} + \text{i}\ket{31})_{AB},
 \end{split}
\end{eqnarray}
\begin{eqnarray}   \label{eq52}
 \begin{split}
    \ket{\psi_{3,3}}_{AB} = \frac{1}{2} (\ket{03} - \text{i}\ket{10} - \ket{21} + \text{i}\ket{32})_{AB}.
 \end{split}
\end{eqnarray}
Then, $\mathcal{S}^4$ transforms the generalized Bell states $\{\ket{\psi_{i,j}}_{AB}\}$ into
\begin{eqnarray}   \label{eq53}
 \begin{split}
    \mathcal{S}^4 \ket{\mathcal{H}_0}^c  \ket{\psi_{0,0}}_{AB} = \ket{\mathcal{H}_0}^c \otimes \ket{\psi_{0,1}}_{AB},
 \end{split}
\end{eqnarray}
\begin{eqnarray}   \label{eq54}
 \begin{split}
    \mathcal{S}^4 \ket{\mathcal{H}_0}^c  \ket{\psi_{0,1}}_{AB} = \ket{\mathcal{H}_0}^c \otimes \ket{\psi_{0,2}}_{AB},
 \end{split}
\end{eqnarray}
\begin{eqnarray}   \label{eq55}
 \begin{split}
    \mathcal{S}^4 \ket{\mathcal{H}_0}^c  \ket{\psi_{0,2}}_{AB} = \ket{\mathcal{H}_0}^c \otimes \ket{\psi_{0,3}}_{AB},
 \end{split}
\end{eqnarray}
\begin{eqnarray}   \label{eq56}
 \begin{split}
    \mathcal{S}^4 \ket{\mathcal{H}_0}^c  \ket{\psi_{0,3}}_{AB} = \ket{\mathcal{H}_0}^c \otimes \ket{\psi_{0,0}}_{AB},
 \end{split}
\end{eqnarray}
\begin{eqnarray}   \label{eq57}
 \begin{split}
    \mathcal{S}^4 \ket{\mathcal{H}_0}^c  \ket{\psi_{1,0}}_{AB} = \ket{\mathcal{H}_3}^c \otimes \ket{\psi_{1,1}}_{AB},
 \end{split}
\end{eqnarray}
\begin{eqnarray}   \label{eq58}
 \begin{split}
    \mathcal{S}^4 \ket{\mathcal{H}_0}^c  \ket{\psi_{1,1}}_{AB} = \ket{\mathcal{H}_3}^c \otimes \ket{\psi_{1,2}}_{AB},
 \end{split}
\end{eqnarray}
\begin{eqnarray}   \label{eq59}
 \begin{split}
    \mathcal{S}^4 \ket{\mathcal{H}_0}^c  \ket{\psi_{1,2}}_{AB} = \ket{\mathcal{H}_3}^c \otimes \ket{\psi_{1,3}}_{AB},
 \end{split}
\end{eqnarray}
\begin{eqnarray}   \label{eq60}
 \begin{split}
    \mathcal{S}^4 \ket{\mathcal{H}_0}^c  \ket{\psi_{1,3}}_{AB} = \ket{\mathcal{H}_3}^c \otimes \ket{\psi_{1,0}}_{AB},
 \end{split}
\end{eqnarray}
\begin{eqnarray}   \label{eq61}
 \begin{split}
    \mathcal{S}^4 \ket{\mathcal{H}_0}^c  \ket{\psi_{2,0}}_{AB} = \ket{\mathcal{H}_2}^c \otimes \ket{\psi_{2,1}}_{AB},
 \end{split}
\end{eqnarray}
\begin{eqnarray}   \label{eq62}
 \begin{split}
    \mathcal{S}^4 \ket{\mathcal{H}_0}^c \ket{\psi_{2,1}}_{AB} = \ket{\mathcal{H}_2}^c \otimes \ket{\psi_{2,2}}_{AB},
 \end{split}
\end{eqnarray}
\begin{eqnarray}   \label{eq63}
 \begin{split}
    \mathcal{S}^4 \ket{\mathcal{H}_0}^c  \ket{\psi_{2,2}}_{AB} = \ket{\mathcal{H}_2}^c \otimes \ket{\psi_{2,3}}_{AB},
 \end{split}
\end{eqnarray}
\begin{eqnarray}   \label{eq64}
 \begin{split}
    \mathcal{S}^4 \ket{\mathcal{H}_0}^c  \ket{\psi_{2,3}}_{AB} = \ket{\mathcal{H}_2}^c \otimes \ket{\psi_{2,0}}_{AB},
 \end{split}
\end{eqnarray}
\begin{eqnarray}   \label{eq65}
 \begin{split}
    \mathcal{S}^4 \ket{\mathcal{H}_0}^c  \ket{\psi_{3,0}}_{AB} = \ket{\mathcal{H}_1}^c \otimes \ket{\psi_{3,1}}_{AB},
 \end{split}
\end{eqnarray}
\begin{eqnarray}   \label{eq66}
 \begin{split}
    \mathcal{S}^4 \ket{\mathcal{H}_0}^c \ket{\psi_{3,1}}_{AB} = \ket{\mathcal{H}_1}^c \otimes \ket{\psi_{3,2}}_{AB},
 \end{split}
\end{eqnarray}
\begin{eqnarray}   \label{eq67}
 \begin{split}
    \mathcal{S}^4 \ket{\mathcal{H}_0}^c \ket{\psi_{3,2}}_{AB} = \ket{\mathcal{H}_1}^c \otimes \ket{\psi_{3,3}}_{AB},
 \end{split}
\end{eqnarray}
\begin{eqnarray}   \label{eq68}
 \begin{split}
   \mathcal{S}^4  \ket{\mathcal{H}_0}^c \ket{\psi_{3,3}}_{AB} = \ket{\mathcal{H}_1}^c \otimes \ket{\psi_{3,0}}_{AB},
 \end{split}
\end{eqnarray}

Based on Eq. \eqref{eq37} - Eq. \eqref{eq68}, one can see that the sixteen generalized Bell states can be completely heralded by the outcomes of the quantum 4-switch $\mathcal{S}^4$, Alice, and Bob, see Table \ref{Tab2}.

\begin{table*}[htbp]
\centering
\caption{The correspondences between the outcome of the quantum 4-switch $\mathcal{S}^4$, ququart information, and the state transition for discriminating sixteen Bell states.} \label{Tab2}

\begin{tabular}{ cccc } 
\hline\hline
 Outcome of 4-switch $\mathcal{S}^4$ & Ququart information of ($j_A, j_B$) &  Output Bell state   &  Input Bell state \\

\hline

\multirow{4}{*}{ $\ket{\mathcal{H}_0}^c$} 
                                     &  ${j_B} = ({j_A} + 1)\bmod 4$  & $\ket{\psi_{0,1}}_{AB}$ & $\ket{\psi_{0,0}}_{AB}$ \\
                                     &  ${j_B} = ({j_A} + 2)\bmod 4$  & $\ket{\psi_{0,2}}_{AB}$ & $\ket{\psi_{0,1}}_{AB}$ \\
                                     &  ${j_B} = ({j_A} +3)\bmod 4$   & $\ket{\psi_{0,3}}_{AB}$ & $\ket{\psi_{0,2}}_{AB}$ \\
                                     &  $j_A = j_B$                   & $\ket{\psi_{0,0}}_{AB}$ & $\ket{\psi_{0,3}}_{AB}$ \\
\addlinespace[0.8em]

\multirow{4}{*}{ $\ket{\mathcal{H}_3}^c$}
                                     &  ${j_B} = ({j_A} + 1)\bmod 4$  & $\ket{\psi_{1,1}}_{AB}$ & $\ket{\psi_{1,0}}_{AB}$ \\
                                     &  ${j_B} = ({j_A} + 2)\bmod 4$  & $\ket{\psi_{1,2}}_{AB}$ & $\ket{\psi_{1,1}}_{AB}$ \\
                                     &  ${j_B} = ({j_A} +3)\bmod 4$   & $\ket{\psi_{1,3}}_{AB}$ & $\ket{\psi_{1,2}}_{AB}$ \\
                                     &  $j_A = j_B$                   & $\ket{\psi_{1,0}}_{AB}$ & $\ket{\psi_{1,3}}_{AB}$ \\
\addlinespace[0.8em]

\multirow{4}{*}{ $\ket{\mathcal{H}_2}^c$} 
                                    &  ${j_B} = ({j_A} + 1)\bmod 4$  & $\ket{\psi_{2,1}}_{AB}$ & $\ket{\psi_{2,0}}_{AB}$ \\
                                    &  ${j_B} = ({j_A} + 2)\bmod 4$  & $\ket{\psi_{2,2}}_{AB}$ & $\ket{\psi_{2,1}}_{AB}$ \\
                                    &  ${j_B} = ({j_A} + 3)\bmod 4$  & $\ket{\psi_{2,3}}_{AB}$ & $\ket{\psi_{2,2}}_{AB}$ \\
                                    &  $j_A = j_B$                   & $\ket{\psi_{2,0}}_{AB}$ & $\ket{\psi_{2,3}}_{AB}$ \\
\addlinespace[0.8em]

\multirow{4}{*}{ $\ket{\mathcal{H}_1}^c$} 
                                    & ${j_B} = ({j_A} + 1)\bmod 4$  & $\ket{\psi_{3,1}}_{AB}$ & $\ket{\psi_{3,0}}_{AB}$ \\
                                    & ${j_B} = ({j_A} + 2)\bmod 4$  & $\ket{\psi_{3,2}}_{AB}$ & $\ket{\psi_{3,1}}_{AB}$ \\ 
                                    & ${j_B} = ({j_A} + 3)\bmod 4$  & $\ket{\psi_{3,3}}_{AB}$ & $\ket{\psi_{3,2}}_{AB}$ \\ 
                                    & $j_A = j_B$                   & $\ket{\psi_{3,0}}_{AB}$ & $\ket{\psi_{3,3}}_{AB}$ \\

\hline\hline
\end{tabular}
\end{table*}

\subsection{Deterministic Bell-state analyzer in d-dimensional system via ICO d-switch}\label{sec24}

Our approach can also be extended to arbitrary $d$-dimensional (qudit, i.e., $d>4$ system) scenario.
The ICO $d$-switch $\mathcal{S}^d$ operation can be written as
\begin{eqnarray}\label{eq69Cite}
\begin{split}
   \mathcal{S}^d |\mathcal{D}_0\rangle^c |\varphi_1\rangle^t_A |\varphi_2\rangle_B = &\;
   \frac{1}{\sqrt{d}} (|0\rangle^c   \otimes U_1^d |\varphi_1\rangle^t_A \otimes V_1^d |\varphi_2\rangle^t_B    
                     + |1\rangle^c   \otimes U_2^d |\varphi_1\rangle^t_A \otimes V_2^d |\varphi_2\rangle^t_B  \\&\quad + \cdots 
                     + |d-1\rangle^c \otimes U_d^d |\varphi_1\rangle^t_A \otimes V_d^d |\varphi_2\rangle^t_B).
\end{split}
\end{eqnarray}
Here and henceforth,
\begin{eqnarray}\label{eq70}
|\mathcal{D}_i\rangle^c = \frac{1}{\sqrt{d}} (\omega^{0     \cdot i}|0\rangle^c 
                                            + \omega^{1     \cdot i}|1\rangle^c +  \cdots  
                                            + \omega^{(d-1) \cdot i}|d-1\rangle^c), \;\;\; i=0,1,\cdots,(d-1).
\end{eqnarray}

If the embedded local qudit operations $U_k^d$ and $V_k^d$ with $k=1,2,\cdots, d$ within $\mathcal{S}^d$ are adjusted as
\begin{eqnarray}\label{eq71}
  U_k^d = (U_{\text{shift}}^d)^{k-1},\quad
  V_k^d = (U_{\text{shift}}^d)^k,
\end{eqnarray}
where 
\begin{eqnarray}\label{eq72}
{U_{\text{shift}}^d} = 
    \begin{pmatrix}
    0      & 0      & \dots  & 0      & 1      \\
    1      & 0      & \dots  & 0      & 0      \\
    0      & 1      & \ddots & \vdots & \vdots \\
    \vdots & \ddots & \ddots & 0      & 0      \\
    0      & \dots  & 0      & 1      & 0
\end{pmatrix}_{d \times d}.
\end{eqnarray}
Then, the evolutions of a set of generalized Bell states $\{|\varphi_{i,j}\rangle\}$ in $d$-dimensional system induced by $\mathcal{S}^d$ can be written as
\begin{eqnarray}\label{eq73}
 \mathcal{S}^d |\mathcal{D}_0\rangle^c|\varphi_{i,j}\rangle_{AB} = |\mathcal{D}_{(d-i)\bmod{d}}\rangle^c \otimes |\varphi_{i,(j+1)\bmod{d}}\rangle_{AB},
\end{eqnarray}
where
\begin{eqnarray}\label{eq74}
|\varphi_{i,j}\rangle_{AB} = \frac{1}{\sqrt{d}} \sum_{k=0}^{d-1} \omega^{ik} |k\rangle_{A} |(k+j)\bmod{d}  \rangle_{B},\;\;
                        i, j \in \{\ 0, 1, \dots, (d-1)\} \;\text{and}\; \omega=e^{\frac{2\pi \text{i}}{d}}.
\end{eqnarray}

Based on Eq. \eqref{eq73}, one can see that regardless of the system dimension $d$, the $d^2$ generalized Bell states can be divided into $d$ distinguishable groups,  $\{|\varphi_{i,0}\rangle_{AB}, |\varphi_{i,1}\rangle_{AB}, \cdots, |\varphi_{i,(d-1)}\rangle_{AB}\}$ corresponding to $|\mathcal{D}_{(d-i)\bmod{d}}\rangle^c$, $i=0,1,\cdots, (d-1)$. 
Subsequently, the particles held by Alice and Bob are measured in the basis $\{\ket{0},\ket{1},\cdots,\ket{d-1}\}$, respectively. Based on the measured results, the states in each group can be distinguished from each other perfectly.

We further find that, after the $d$-switch process, the indefinite causal structure of the quantum switch does not get consumed.
Hence the nondestructive BSA can be achieved by iterating the $d$-switch process for two rounds.

\section{Physical implementation of BSA in $d$-dimensional system via gravitational ICO switch} \label{sec3}

The ICO switch $\mathcal{S}^d$ is the key element of our high-dimensional BSA protocols. It is known that space-time physics \cite{space-time1}, causal modeling \cite{causal-model}, time-delocalization \cite{time-delocalization}, and time-direction \cite{ICO-confirm3} are the viable candidates for quantum switch.
Inspired by the gravitational ICO quantum 2-switch  proposed in Ref. \cite{gravity-BSA}, we will introduce the working principle of gravitational (space-time) ICO quantum $d$-switch $\mathcal{S}^d$ for implementing our high-dimensional BSA protocols.

\subsection{The gravitational ICO 3-switch  $\mathcal{S}^3$ for BSA in 3-dimensional system}\label{sec31}

We now detail how the ICO 3-switch $\mathcal{S}^3$ mechanism (shown in Sec. \ref{sec22}) can be implemented utilizing the gravitational effects induced by the matter-energy distribution.

Consider two initially synchronized clocks (agents), Alice and Bob, placed in the gravitational field of a  massive object $M$. $r_A$ ($r_B$) is the spatial distance between the mass and Alice (Bob).
Consider three events: Event A, defined by Alice’s clock at her local time $\tau$. 
Events $B_1$ and $B_2$, defined by Bob’s clock at his local times $\tau$ and $\tau_1$, respectively.
Here $\tau_1 > \tau$, that is, $B_1$ and $B_2$ are always in the fixed order, i.e., $B_1$ is always in the causal past of $B_2$ ($B_1 \rightarrow B_2$).

\begin{figure*} 
\includegraphics[width=14 cm,angle=0]{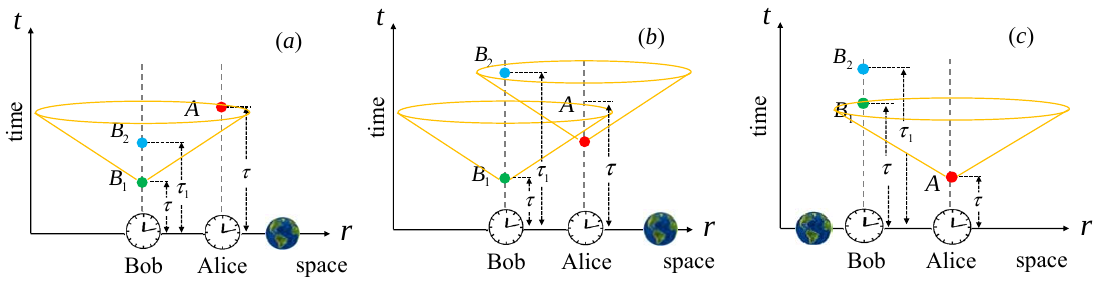}
\caption{Quantum control of temporal order for constructing gravitational ICO  3-switch $\mathcal{S}^3$.
  (a) Alice's clock is located near the mass object with $r_A=R_1$     and $r_B=R_1+h$. Events $B_1$, $B_2$, and $A$  will occur in succession, i.e., $M_{B_1 \rightarrow B_2 \rightarrow A}$.   
  (b) Alice's clock is located near the mass object with $r_A=R_2>R_1$ and $r_B=R_2+h$. Events $B_1$, $A$,   and $B_2$ will occur in succession, i.e., $M_{B_1 \rightarrow A \rightarrow B_2}$.
  (c) Bob's clock is located near the mass object   with $r_B=R_1$     and $r_A=R_1+h$. Events $A$, $B_1$,   and $B_2$ will occur in succession, i.e., $M_{A \rightarrow B_1 \rightarrow B_2}$.
   $r_A$ ($r_B$) is the spatial distance between the mass and Alice (Bob).}
 \label{Fig1}
\end{figure*}

\begin{table*}[htbp]
\centering
\caption{The correspondence between the signal and operation in each configuration in 3-dimensional quantum system.}\label{tab3} 
\begin{tabular}{cccc} 
\hline \hline
        Agent                       &                Configuration         &          Scenario                                &
         Operation \\
\hline
\multirow{3}{*}{Alice}    & $\ket{M_{B_1 \rightarrow B_2 \rightarrow A}}$  &  Alice receives Bob's signals $b_1$ and $b_2$  before $t_A=\tau$      & $U^3_{A|B_2}$  \\
 
                          & $\ket{M_{B_1 \rightarrow A \rightarrow B_2}}$  &  Alice receives Bob's signal $b_1$  only  before $t_A=\tau$           & $U^3_{A|B_1}$ \\
  
                          & $\ket{M_{A \rightarrow B_1 \rightarrow B_2}}$  &  No signal received  before $t_A=\tau$                                & $U^3_{A}$ \\
\addlinespace[0.8em]

\multirow{3}{*}{Bob}     & $\ket{M_{B_1 \rightarrow B_2 \rightarrow A}}$   &  No signal received  before $t_B=\tau_1$                              & $ U^3_{B_2}  U^3_{B_1}$ \\
 
                         & $\ket{M_{B_1 \rightarrow A \rightarrow B_2}}$   &  Bob only receives Alice's signal $a$  received before $t_B=\tau_1$   & $U^3_{B_2|A} U^3_{B_1}$ \\

                         & $\ket{M_{A \rightarrow B_1 \rightarrow B_2}}$   &  Bob receives Alice's signals $a$ before $t_B=\tau$                   & $U^3_{B_2|A} U^3_{B_1|A}$ \\
\hline\hline
\end{tabular}
\end{table*}

In general relativity, time passes more slowly closer to the massive object due to gravitational time dilation. 
Note that a physical system can only be transferred from the past to the future, but not the other way around.

As shown in Table \ref{tab3}, depending on the geometry of the mass object, there are three types of causal orders (after local operation is completed, the agent will send the light signals to each other):

(1) Configuration $M_{B_1 \rightarrow B_2 \rightarrow A}$. 
As shown in Fig. \ref{Fig1}(a), Alice's clock is located near the mass object ($r_A=R_1$, $r_B=R_1+h$).
The gravitational time dilation results in the causal order $B_1 \rightarrow B_2 \rightarrow A$. 
Consequently, event $A$ lies in the future light cone of event $B_2$. 
The corresponding unitary operations {performed} at the space-time events $B_1$, $B_2$, and $A$ are $U^3_{B_1}$, $U^3_{B_2}$, and $U^3_{A|B_2}$, respectively.

(2) Configuration $M_{B_1 \rightarrow A \rightarrow B_2}$. 
For this configuration, shown in Fig. \ref{Fig1}(b), the clocks are positioned such that $r_A=R_2>R_1$ and $r_B=R_2+h$. This arrangement leads to the causal order $B_1 \rightarrow A \rightarrow B_2$. 
Therefore, event $A$ lies in the future light cone of event $B_1$, and event $B_2$ lies in the future light cone of event $A$. The unitary operations {performed} at the spacetime events $B_1$, $A$, and $B_2$ are $U^3_{B_1}$, $U^3_{A|B_1}$, and $U^3_{B_2|A}$, respectively.

(3) Configuration $M_{A \rightarrow B_1 \rightarrow B_2}$. 
In the third configuration, shown in Fig. \ref{Fig1}(c), Bob's clock is closer to the mass object ($r_B=R_1$, $r_A=R_1+h$), resulting in the causal order $A \rightarrow B_1 \rightarrow B_2$. This means both events $B_1$ and $B_2$ lie in the future light cone of event $A$. The corresponding unitary operations {performed} at the spacetime events $A$, $B_1$, and $B_2$ are $U^3_{A}$, $U^3_{B_1|A}$, and $U^3_{B_2|A}$, respectively.

\begin{figure} [htpb]
\begin{center}
\includegraphics[width=5.5 cm,angle=0]{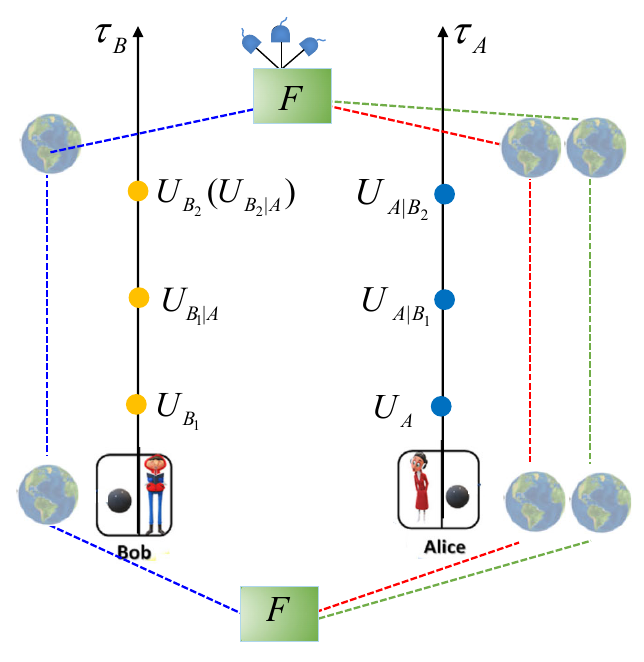}
\caption{The gravitational ICO  3-switch $\mathcal{S}^3$.  $U^3_{A}$, $U^3_{B_1}$, $U^3_{B_2}$, $U^3_{B_1|A}$,  $U^3_{B_2|A}$, $U^3_{A|B_1}$, $ U^3_{B_1|A}$, and $U^3_{B_2|A}$ are the local single-qutrit unitary operations.} \label{Fig2}
\end{center}
\end{figure}

Note that the causal structure of the space-time, as shown in Fig. \ref{Fig1}, is definite.
Fortunately, the quantum superposition of three distinct locations
\begin{eqnarray}   \label{eq75}
\begin{split}
  |\mathcal{F}_0\rangle^c =\; & \frac{1}{\sqrt{3}} (|M_{B_1 \rightarrow B_2 \rightarrow A}\rangle
                                                  + |M_{B_1 \rightarrow A \rightarrow B_2}\rangle
                                                  + |M_{A \rightarrow B_1 \rightarrow B_2}\rangle)
\end{split}
\end{eqnarray}
is the result of a process wherein the order of operation performed on the target system is determined by the position of the mass object.
The operation of this gravitational ICO  3-switch $\mathcal{S}^3$ depicted in Fig. \ref{Fig2} can be written as
\begin{eqnarray} \label{eq76}
\begin{split}
\mathcal{S}^3|\Psi\rangle_A^t |\Psi\rangle_B^t |\mathcal{F}_0\rangle^c = \; & \frac{1}{\sqrt{3}}
  \big(|M_{B_1 \rightarrow B_2 \rightarrow A}   \rangle (U^3_{A|B_2}) |\Psi\rangle_A^t  \otimes  (U^3_{B_2}   U^3_{B_1})    |\Psi\rangle_B^t \\
    &+ |M_{B_1 \rightarrow A   \rightarrow B_2} \rangle (U^3_{A|B_1}) |\Psi\rangle_A^t  \otimes  (U^3_{B_2|A} U^3_{B_1})   |\Psi\rangle_B^t \\
    &+ |M_{A   \rightarrow B_1 \rightarrow B_2} \rangle (U^3_{A})     |\Psi\rangle_A^t  \otimes  (U^3_{B_2|A} U^3_{B_1|A}) |\Psi\rangle_B^t \big).
\end{split}
\end{eqnarray}
The state $|\mathcal{F}_0\rangle^c$, given in Eq.~\eqref{eq26} - Eq.~\eqref{eq28}, can be identified as the control qutrit of the gravitational ICO  3-switch as it coherently governs the order of operations acting on the target system.

The correspondences between Eq. \eqref{eq4Cite} and Eq. \eqref{eq76} are
\begin{eqnarray} \label{eq77}
\begin{split}
  |M_{B_1 \rightarrow B_2 \rightarrow A}  \rangle \equiv |0\rangle^c,\;\;
  |M_{B_1 \rightarrow A   \rightarrow B_2}\rangle \equiv |1\rangle^c,\;\;
  |M_{A   \rightarrow B_1 \rightarrow B_2}\rangle \equiv |2\rangle^c,
\end{split}
\end{eqnarray}
\begin{eqnarray} \label{eq78}
\begin{split}
   U^3_{A|B_2} \equiv U_1^3, \;\;
   U^3_{A|B_1} \equiv U_2^3, \;\;
   U^3_{A}     \equiv U_3^3,
\end{split}
\end{eqnarray}
\begin{eqnarray} \label{eq79}
\begin{split}
   U^3_{B_2}   U^3_{B_1}  \equiv V_1^3,\;\;        
   U^3_{B_2|A} U^3_{B_1}  \equiv V_2^3,\;\; 
   U^3_{B_2|A} U^3_{B_1|A}\equiv V_3^3,
\end{split}
\end{eqnarray}
where
\begin{eqnarray}   \label{eq80}
\begin{split}
    U^3_{A|B_1}= U^3_{\text{shift}},\quad
    U^3_{A} = U^3_{B_1}= U^3_{B_2} = (U^3_{\text{shift}})^{2},\quad
    U^3_{B_1|A} = U^3_{B_2|A}=U^3_{A|B_2}= I_3.
\end{split}
\end{eqnarray}

\subsection{The gravitational ICO  4-switch $\mathcal{S}^4$ for BSA in 4-dimensional system}\label{sec32}

\begin{figure*}
\begin{center}
\includegraphics[width=12 cm,angle=0]{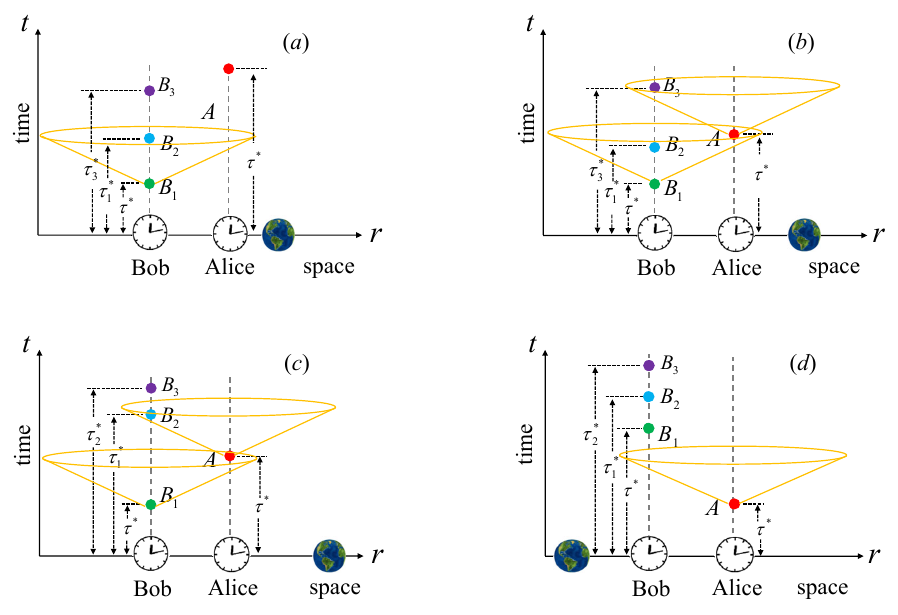}
\caption{Quantum control of temporal order for gravitational ICO  4-switch $\mathcal{S}^4$. 
       (a) The space-time geometry is $M_{B_1 \rightarrow B_2 \rightarrow B_3 \rightarrow A}$. 
       (b) The space-time geometry is $M_{B_1 \rightarrow B_2 \rightarrow A \rightarrow B_3}$. 
       (c) The space-time geometry is $M_{B_1 \rightarrow A \rightarrow B_2 \rightarrow B_3}$. 
       (d) The space-time geometry is $M_{A \rightarrow B_1 \rightarrow B_2\rightarrow B_3}$.} \label{Fig3}
\end{center}
\end{figure*}

As shown in Fig. \ref{Fig3}, we consider four events: Event A, defined by Alice's clock at time $\tau^*$. 
Events $B_1$, $B_2$, and $B_3$, defined by Bob's clock at times $\tau^*$, $\tau_1^*$, and $\tau_2^*$, respectively.
Here $\tau_2^* > \tau_1^* > \tau^*$.
As shown in Table \ref{tab4}, the four types of causal orders are:

(1) Configuration $M_{B_1 \rightarrow B_2 \rightarrow B_3 \rightarrow A}$.
    As shown in Fig. \ref{Fig3}(a), the corresponding order of unitary operations performed by two agents is $U^4_{B_1}$, $U^4_{B_2}$, $U^4_{B_3}$, and $U^4_{A|B_3}$.

(2) Configuration $M_{B_1 \rightarrow B_2 \rightarrow A \rightarrow B_3}$.
    As shown in Fig. \ref{Fig3}(b), the corresponding order of unitary operations performed by two agents is $U^4_{B_1}$, $U^4_{B_2}$, $U^4_{A|B_1}$, and $U^4_{B_3|A}$.

(3) Configuration $M_{B_1 \rightarrow A \rightarrow B_2 \rightarrow B_3}$.
    As shown in Fig. \ref{Fig3}(c), the corresponding order of unitary operations performed by two agents is $U^4_{B_1}$, $U^4_{A|B_1}$, $U^4_{B_2|A}$, and $U^4_{B_3|A}$.

(4) Configuration $M_{A \rightarrow B_1 \rightarrow B_2 \rightarrow B_3 }$.
    As shown in Fig. \ref{Fig3}(d), the corresponding order of unitary operations performed by two agents is $U^4_A$, $U^4_{B_1|A}$, $U^4_{B_2|A}$, and $U^4_{B_3|A}$.

\begin{figure}
\begin{center}
\includegraphics[width=5 cm,angle=0]{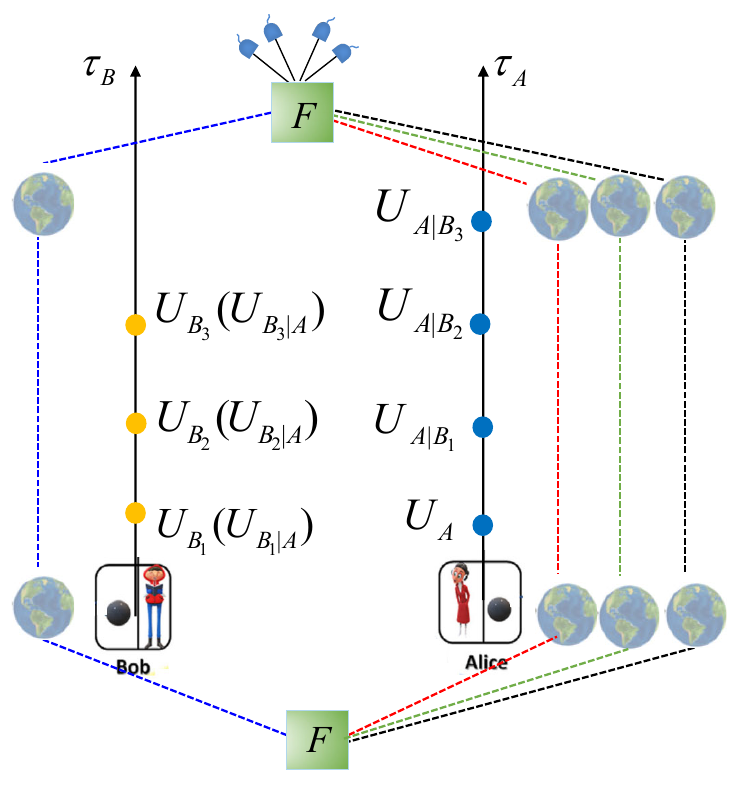}
\caption{The gravitational ICO  4-switch  $\mathcal{S}^4$.  
        $U^4_{A}$, $U^4_{B_1}$, $U^4_{B_2}$, $U^4_{B_3}$, $U^4_{B_1|A}$, $U^4_{B_2|A}$, $U^4_{B_3|A}$, $U^4_{A|B_1}$, $U^4_{A|B_2}$, and $U^4_{A|B_3}$ are the embedded local single-ququart unitary operations.} \label{Fig4}
\end{center}
\end{figure}

Therefore, the operation of the gravitational ICO  4-switch  $\mathcal{S}^4$, depicted in Fig. \ref{Fig4},  can be expressed as
\begin{eqnarray} \label{eq81}  
\begin{split}
\mathcal{S}^4|\psi\rangle_A^t |\psi\rangle_B^t |\mathcal{H}_0\rangle^c =\; &
\frac{1}{2}(|M_{B_1 \rightarrow B_2 \rightarrow B_3 \rightarrow A}  \rangle (U^4_{A|B_3}) |\psi\rangle_A^t \otimes (U^4_{B_3}   U^4_{B_2}   U^4_{B_1})      |\psi\rangle_B^t\\&
          + |M_{B_1 \rightarrow B_2 \rightarrow A   \rightarrow B_3}\rangle (U^4_{A|B_2}) |\psi\rangle_A^t \otimes (U^4_{B_3|A} U^4_{B_2}   U^4_{B_1})    |\psi\rangle_B^t\\&
          + |M_{B_1 \rightarrow A   \rightarrow B_2 \rightarrow B_3}\rangle (U^4_{A|B_1}) |\psi\rangle_A^t  \otimes(U^4_{B_3|A} U^4_{B_2|A} U^4_{B_1})  |\psi\rangle_B^t\\&
          + |M_{A   \rightarrow B_1 \rightarrow B_2 \rightarrow B_3}\rangle (U^4_{A})     |\psi\rangle_A^t \otimes (U^4_{B_3|A} U^4_{B_2|A} U^4_{B_1|A})|\psi\rangle_B^t).
\end{split}\label{eq_a1}
\end{eqnarray}
The correspondences between Eq. \eqref{eq29Cite} and Eq. \eqref{eq81} are
\begin{eqnarray} \label{eq82}
\begin{split}
& |M_{B_1 \rightarrow B_2 \rightarrow B_3 \rightarrow A}  \rangle \equiv |0\rangle^c,\;\;
  |M_{B_1 \rightarrow B_2 \rightarrow A   \rightarrow B_3}\rangle \equiv |1\rangle^c,\\
& |M_{B_1 \rightarrow A   \rightarrow B_2 \rightarrow B_3}\rangle \equiv |2\rangle^c,\;\;
  |M_{A   \rightarrow B_1 \rightarrow B_2 \rightarrow B_3}\rangle \equiv |3\rangle^c,
\end{split}
\end{eqnarray}
\begin{eqnarray} \label{eq83}
\begin{split}
   U^4_{A|B_3} \equiv U_1^4,\;\;
   U^4_{A|B_2} \equiv U_2^4,\;\;
   U^4_{A|B_1} \equiv U_3^4,\;\; 
   U^4_{A}     \equiv U_4^4,
\end{split}
\end{eqnarray}
\begin{eqnarray} \label{eq84}
\begin{split}
   U^4_{B_3}   U^4_{B_2}   U^4_{B_1}   \equiv V_1^4,\;        
   U^4_{B_3|A} U^4_{B_2}   U^4_{B_1}   \equiv V_2^4, \;  
   U^4_{B_3|A} U^4_{B_2|A} U^4_{B_1}   \equiv V_3^4,\;
   U^4_{B_3|A} U^4_{B_2|A} U^4_{B_1|A} \equiv V_4^4,
\end{split}
\end{eqnarray}
where
\begin{eqnarray}   \label{eq85}
\begin{split}
    U^4_{A|B_2}= U^4_{\text{shift}},\quad
    U^4_{A|B_1} = (U^4_{\text{shift}})^{2},
\end{split}
\end{eqnarray}
\begin{eqnarray}   \label{eq86}
\begin{split}
   U^4_{B_1} = U^4_{B_2}= U^4_{B_2} = U^4_{A} = (U^4_{\text{shift}})^{3},
     \end{split}
\end{eqnarray}
\begin{eqnarray}   \label{eq87}
\begin{split}
   U^4_{B_1|A} = U^4_{B_2|A}= U^4_{B_3|A} = U^4_{A|B_3} = I_4.\quad
\end{split}
\end{eqnarray}

\begin{table*}[htbp]
  \centering
  \caption{The correspondences between the signal and operation in each configuration in 4-dimensional quantum system.} \label{tab4}
  \begin{tabular}{cccc}
    \hline\hline
    Agent                    &                Configuration         &          Scenario                                &       Operation \\
    \hline
    \multirow{8}{*}{Alice} 
                             & $\ket{M_{B_1 \rightarrow B_2 \rightarrow B_3 \rightarrow A}}$ & Received Bob's $b_1$, $b_2$ and $b_3$ signals before $t_A=\tau^{*}$ & $U^4_{A|B_3}$ \\
&&&\\
                             & $\ket{M_{B_1 \rightarrow B_2 \rightarrow A \rightarrow B_3}}$ & Received Bob's $b_1$, $b_2$ signals before $t_A=\tau^{*}$           & $U^4_{A|B_2}$ \\
&&&\\
                             & $\ket{M_{B_1 \rightarrow A \rightarrow B_2 \rightarrow B_3}}$ & Received Bob's $b_1$ signal before $t_A=\tau^{*}$                   & $U^4_{A|B_1}$ \\
&&&\\
                             & $\ket{M_{A \rightarrow B_1 \rightarrow B_2 \rightarrow B_3}}$ & No signal received before $t_A=\tau^{*}$                            & $U^4_{A}$ \\
\addlinespace[2.4em]

    \multirow{8}{*}{Bob}   
                             & $\ket{M_{B_1 \rightarrow B_2 \rightarrow B_3 \rightarrow A}}$ & No signal received before $t_B=\tau_2^{*}$    &  $U^4_{B_3} U^4_{B_2} U^4_{B_1}$ \\
&&&\\
                             &  \multirow{2}{*}{$\ket{M_{B_1 \rightarrow B_2 \rightarrow A \rightarrow B_3}}$} & \multirow{1}{*}{ No signal received before $t_B=\tau_1^{*}$,}  
                                                                                                               & \multirow{2}{*}{$U^4_{B_3|A} U^4_{B_2} U^4_{B_1}$}\\

                             & &\multirow{1}{*}{ but  Alice's signals received before $t_B=\tau_2^{*}$}        &  \\

                             &  \multirow{2}{*}{$\ket{M_{B_1 \rightarrow A \rightarrow B_2 \rightarrow B_3}}$} & \multirow{1}{*}{No signal received before $t_B=\tau^{*}$,} 
                                                                                                               & \multirow{2}{*}{$U^4_{B_3|A} U^4_{B_2|A} U^4_{B_1}$ }\\
                             & &\multirow{1}{*}{ but  Alice's signals received before $t_B=\tau_1^{*}$}        & \\
   &&&\\                                 
                             & $\ket{M_{A \rightarrow B_1 \rightarrow B_2 \rightarrow B_3}}$                   & Alice's signals received before $t_B=\tau^{*}$  
                                                                                                               & $U^4_{B_3|A} U^4_{B_2|A} U^4_{B_1|A}$ \\
    \hline\hline
  \end{tabular}
\end{table*}

\subsection{The gravitational ICO $d$-switch $\mathcal{S}^d$ for BSA in $d$-dimensional system}\label{sec33}

\begin{figure} 
\begin{center}
\includegraphics[width=12 cm,angle=0]{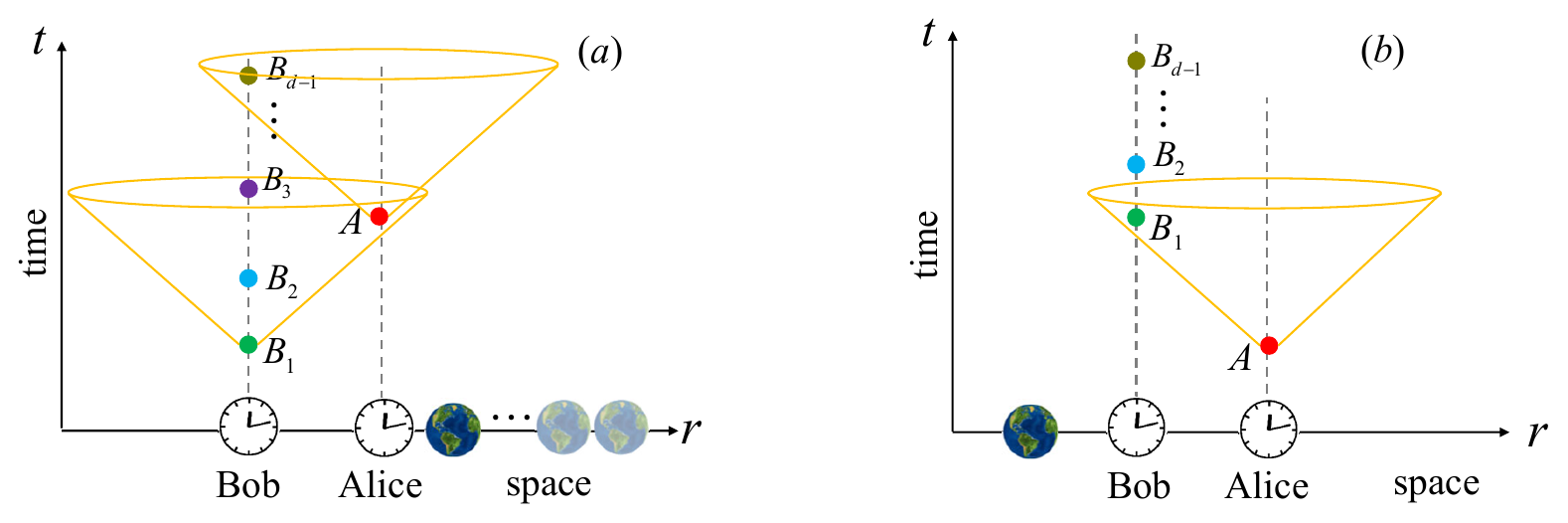}
\caption{Quantum control of temporal order for gravitational ICO  $d$-switch $\mathcal{S}^d$. 
    (a) Alice's clock is located near the mass object. 
    (b) Bob's   clock is located near the mass object.} \label{Fig5}
\end{center}
\end{figure}

As shown in Fig. \ref{Fig5}, consider there are $d$ ($d>4$) events $A$, $B_1$, $B_2$, $\cdots$, $B_{d-1}$.
There are $d$ types of causal orders:

(1) Configuration $M_{B_1 \rightarrow B_2 \rightarrow \dots \rightarrow B_{d-1}\rightarrow A}$.
    The order of unitary operations performed by the two agents is $U^d_{B_1}$, $U^d_{B_2}$, $\cdots$,  $U^d_{B_{d-1}}$, and $U^d_{A|B_{d-1}}$.

(2) Configuration $M_{B_1 \rightarrow B_2 \rightarrow \dots \rightarrow A \rightarrow B_{d-1}}$.
    The order of unitary operations performed by the two agents is $U^d_{B_1}$, $\cdots$, $U^d_{B_{d-2}}$, $U^d_{A|B_{d-2}}$, and $U^d_{B_{d-1}|A}$.

   $\dots$

($d$) Configuration $M_{A \rightarrow B_1 \rightarrow B_2 \rightarrow \dots \rightarrow B_{d-1}}$.
      The order of unitary operations performed by the two agents is $U^d_A$, $U^d_{B_1|A}$, $U^d_{B_2|A}$, $\cdots$, $U^d_{B_{d-1}|A}$.

\begin{figure} 
\begin{center}
\includegraphics[width=6cm,angle=0]{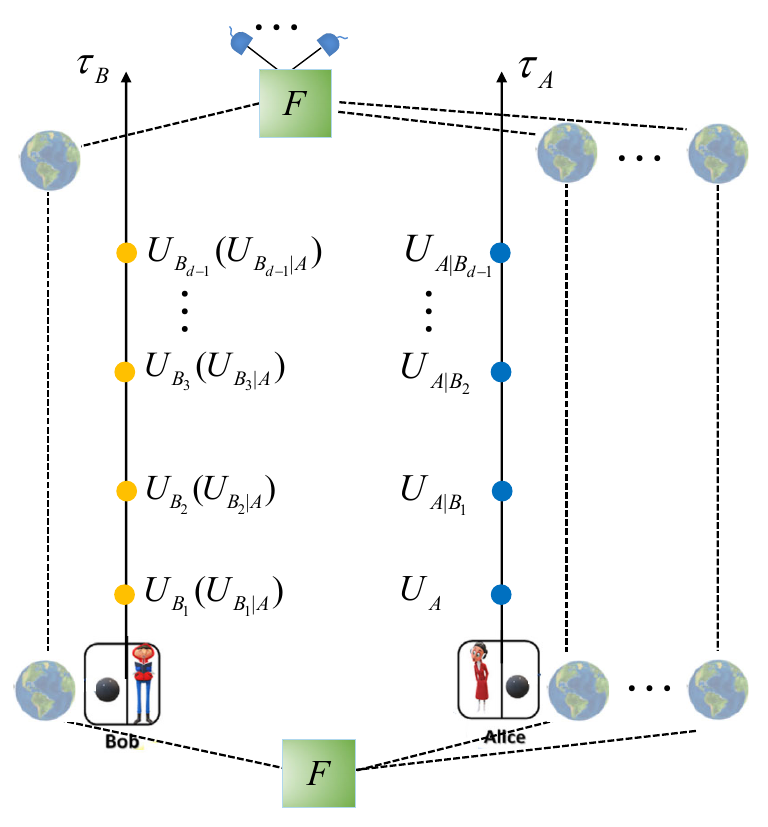}
\caption{The gravitational ICO  $d$-switch $\mathcal{S}^d$.
         $U^d_{A}$, $U^d_{B_1}$, $U^d_{B_2}$, $\cdots$, $U^d_{B_{d-1}}$, $U^d_{B_1|A}$,  $U^d_{B_2|A}$, $\cdots$, 
         $U^d_{{B_{d-1}}|A}$, $U^d_{A|B_1}$, $U_{A|B_2}$, $\cdots$, $U^d_{A|B_{d-1}}$ are the inherent local single-qudit unitary operations.} \label{Fig6}
\end{center}
\end{figure}

The physically allowed mass configuration of gravitational ICO  $d$-switch $\mathcal{S}^d$ can be given by
\begin{eqnarray}  \label{eq88}
\begin{split}
 |\mathcal{D}_0\rangle^c =\;& \frac{1}{\sqrt{d}}
                                      (|M_{B_1 \rightarrow B_2 \rightarrow \dots \rightarrow B_{d-1} \rightarrow A}\rangle  +
                                       |M_{B_1 \rightarrow B_2 \rightarrow \dots \rightarrow A       \rightarrow B_{d-1}}\rangle \\& + \cdots +
                                       |M_{A   \rightarrow B_1 \rightarrow  B_2  \rightarrow \dots   \rightarrow B_{d-1}}\rangle).
\end{split} 
\end{eqnarray}%
Hence the transformation of the $d$-switch $\mathcal{S}^d$ depicted in Fig. \ref{Fig6} can be written as 
\begin{eqnarray}   \label{eq89}
\begin{split}
\mathcal{S}^d  \ket{\varphi}_A^t  \ket{\varphi}_B^t|\mathcal{D}_0\rangle^c = & \frac{1}{\sqrt{d}} 
 (|M_{B_1 \rightarrow B_2 \rightarrow \cdots \rightarrow B_{d-1}\rightarrow A}\rangle (U^d_{A|B_{d-1}}) \ket{\varphi}_A^t \otimes (U^d_{B_{d-1}} U^d_{B_{d-2}} \cdots U^d_{B_1})  
 \ket{\varphi}_B^t \\& 
 +|M_{B_1 \rightarrow B_2 \rightarrow \cdots \rightarrow A \rightarrow B_{d-1}}\rangle (U^d_{A|B_{d-2}})\ket{\varphi}_A^t \otimes (U^d_{B_{d-1}|A} {U^d}_{B_{d-2}} \cdots U^d_{B_1})  \ket{\varphi}_B^t \\&
 +\cdots\\&
 +|M_{A \rightarrow B_1 \rightarrow B_2 \rightarrow \cdots \rightarrow B_{d-1}}\rangle (U^d_A)          \ket{\varphi}_A^t \otimes (U^d_{B_{d-1}|A} U^d_{B_{d-2}|A} \cdots U^d_{B_1|A}) \ket{\varphi}_B^t).
\end{split}
\end{eqnarray}
The correspondences between Eq. \eqref{eq69Cite} and Eq. \eqref{eq89} are
\begin{eqnarray} \label{eq90}
\begin{split}
 &|M_{B_1 \rightarrow B_2 \rightarrow \cdots \rightarrow B_{d-1}\rightarrow A}\rangle   \equiv  |0\rangle^c, \\
 &|M_{B_1 \rightarrow B_2 \rightarrow \cdots \rightarrow A \rightarrow B_{d-1}}\rangle  \equiv  |1\rangle^c,\\& \cdots \\
 &|M_{A   \rightarrow B_1 \rightarrow B_2 \rightarrow \cdots \rightarrow B_{d-1}}\rangle \equiv |d-1\rangle^c,
\end{split}
\end{eqnarray}
\begin{eqnarray} \label{eq91}
\begin{split}
  U^d_{A|B_{d-1}} \equiv U_1^d,\;\;
  U^d_{A|B_{d-2}} \equiv U_2^d,\;\; ...\;,\;\;
  U^d_{A}         \equiv U_d^d, 
\end{split}
\end{eqnarray}
\begin{eqnarray} \label{eq92}
\begin{split}
  U^d_{B_{d-1}}  U^d_{B_{d-2}}   \cdots U^d_{B_1}   \equiv  V_1^d,\;\;   
  U^d_{B_{d-1}|A} U^d_{B_{d-2}}   \cdots U^d_{B_1}   \equiv  V_2^d, \;\;  
  U^d_{B_{d-1}|A} U^d_{B_{d-2}|A} \cdots U^d_{B_1|A} \equiv  V_d^d,
\end{split}
\end{eqnarray}
where
\begin{eqnarray} \label{eq93}
U^d_A = (U^d_{\text{shift}})^{d-1}, \;\;
U^d_{A|B_1} = (U^d_{\text{shift}})^{d-2},\;\;
\cdots,\;\;
U^d_{A|B_k}  = (U^d_{\text{shift}})^{d-1-k},\;\;
\cdots,\;\;
U^d_{A|B_{d-1}} = I_d,
\end{eqnarray}
\begin{eqnarray} \label{eq94}
  U^d_{B_1|A}= U^d_{B_2|A} = \cdots = U^d_{B_{d-1}|A}= I_d,
\end{eqnarray}
\begin{eqnarray} \label{eq95}
U^d_{B_1}  = (U^d_{\text{shift}})^{d-1},
\end{eqnarray}
\begin{eqnarray} \label{eq96}
U^d_{B_2} = U^d_{B_3} = \cdots = U^d_{B_{d-1}} = (U^d_{\text{shift}})^{d - 1}.
\end{eqnarray}

\section{Discussion and conclusion}\label{sec4}

ICO serves as a universal resource for locally implementing various QIP tasks on spatially distributed quantum systems, 
and ICO has mainly focused on 2-dimensional quantum systems.
By exploiting solely single-qubit operations in superposition of causal orders, the task can be completed without  entangled gates or direct interaction between qubits.
While this resource has been applied to high-dimensional tasks such as quantum teleportation  \cite{gravity-BSA}, a complete and deterministic high-dimensional Bell-state analysis (BSA) has remained a largely open problem.

 
In this paper, we proposed complete and deterministic BSA protocols beyond qubit systems via ICO.
We first proposed a ICO quantum 3-switch $\mathcal{S}^3$, and a perfect  $\mathcal{S}^3$-based BSA protocol in 3-dimensional quantum system is proposed.
Subsequently, our approach is extended to 4-dimensional and arbitrary $d$-dimensional scenarios. 
Furthermore, the gravitational ICO 3-switch, ICO 4-switch, and even ICO $d$-switch are designed to complete our BSA schemes. 
The results indicate that, independent of the dimension, perfect one-shot qudit BSA can be achieved via ICO quantum $d$-switch $\mathcal{S}^d$ with $d \geq 3)$.
Maximally entangled states or controlled-unitary gates, which are necessary for previous BSA protocols,  are not required in our scheme.
In standard two-step BSA protocols, the bit and phase information of the Bell states are distinguished sequentially. 
Importantly, in our schemes, the nondestructive BSAs can also be achieved by iterating the $d$-switch process for two rounds. 
Additionally, the intrinsic local unitary operations involved in the gravitational ICO switch are remarkably simple, consisting solely of shift gates.

\data{No new data were created or analysed in this study.}

\ack{This work is supported by the National Natural Science Foundation of China under Grant No. 62371038 and Grant No. 12505028, the Science Research Project of Hebei Education Department under Grant No. QN2025054, the Beijing Natural Science Foundation under Grant No. 4252006, and the Fundamental Research Funds for the Central Universities under Grant No. FRF-TP-19-011A3.}




\providecommand{\newblock}{}

\end{document}